\documentclass[11pt]{article}
\usepackage{bez123,calc,curves,ebezier,epic,eepic,graphicx,multiply,rotating}
\usepackage{algorithmic}
\usepackage{bez123,calc,curves,ebezier,epic,eepic,graphicx,multiply,rotating}
\everymath{\displaystyle}
\usepackage{graphicx}
\usepackage{pst-all}
\usepackage{color}
\usepackage{amssymb}
\usepackage{amsmath}
\newtheorem{prethm}{{\bf Theorem}}

\newtheorem{prelemma}{{\bf Lemma}}

\newtheorem{preex}{{\bf Example}}

\newtheorem{preprop}{{\bf Proposition}}

\newtheorem{precor}{{\bf Corollary}}

\newtheorem{preremark}{{\bf Remark}}

\newtheorem{preprob}{{\bf Problem}}

\newtheorem{predefin}{{\bf Definition}}

\newtheorem{preconj}{{\bf Conjecture}}

\newtheorem{preprobb}{{\bf Problem}}

\newtheorem{prelem}{{\bf Theorem}}

\newtheorem{presolution}{{\bf Solution.}}

\def\newpic#1{}
\def\qed{\ifhmode\unskip\nobreak\fi\quad\ifmmode\Box\else$\Box$\fi}

\title{\vspace{-1cm}\Large\bf\noindent A memory theoretic approach for investigating the roles of language and intuition in mathematical thinking activities}

\author{\large\bf Manouchehr Zaker\footnote{mzaker@iasbs.ac.ir}
\vspace{5mm}\\
    Department of Mathematics,\\
     Institute for Advanced Studies in Basic Sciences,\\
    Zanjan, Iran\\
  }
    \date{}

\begin{document}
\maketitle

\begin{abstract}
\noindent Questions concerning origin of mathematical knowledge and roles of language and intuition (imagery) in mathematical thoughts are long standing and widely debated. By introspection, mathematicians usually have some beliefs regarding these questions. But these beliefs are usually in a big contrast with the recent cognitive theoretic findings concerning mathematics. Contemporary cognitive science opens new approaches to reformulate the fundamental questions concerning mathematics and helps mathematicians break through the Platonic beliefs about the essence and sources of mathematical knowledge. In this article, we introduce and discuss mathematical thinking activities and fundamental processes such as symbolic/formal and visual/spatial ones. Two different aspects of mathematics should be separated in mathematical cognition. One aspect considers mathematics as an explicit crystallized knowledge. The other aspect considers mathematics as an ongoing and transient mental processing. The cognitive processes and corresponding tasks involved in these aspects are different. Ongoing mathematical activities both elementary and advanced, demand working memory resources. Using dual-task techniques, we design some pilot experiments to differentiate the symbolic/formal and visual/spatial processes. Using this memory theoretic approach, we explain the crucial roles of language-based processes such as verbal articulation and instructive speech and also visuo-spatial intuition such as spatial imagery and mental movement in various aspects of mathematics.
\end{abstract}



\section{Introduction}

\noindent What are the roles of language and intuition for mathematical knowledge? Individual opinions and thought schools in philosophy of mathematics are often distinguished by their answer to this controversial question. Logicians such as Bertrand Russell believes (in 1903) that ``The fact that all Mathematics is Symbolic Logic is one of the greatest discoveries of our age; and when this fact has been established, the remainder of the principles of mathematics consists in the analysis of Symbolic Logic itself." \cite{R}. Russell gives no credit to intuition in mathematical thoughts. Formalists such as David Hilbert believe that mathematics is essentially a production of language and symbolic processing and manipulations of mathematical objects. In contrast to formalists, are intuitionists and propounders of mathematical intuition which emphasize on faculties such as imagery, visualization and creativity. Philosophers such as Kant \cite{K} and many mathematicians from Henri Poincar\'{e} \cite{P} to Solomon Feferman \cite{F} insisted on the vital and irreplaceable role of intuition in mathematical knowledge. The dispute between language and intuition has been remained unresolved in philosophy and epistemology of mathematics. Contemporary cognitive science opens new approaches to reformulate the fundamental questions concerning mathematics and helps mathematicians break through the Platonic beliefs about the essence of mathematical thoughts. It is then illuminating and instructive for mathematicians to learn about the main findings of cognitive scientific methodologies addressing the fundamental questions of mathematics. The investigation of roles of language and imagery (intuition) is the research subject of a few disciplines in contemporary cognitive science and the results are striking. Developmental psychology such as Piagetian tradition, investigates the roles of cognitive growth of children in the development of their mathematical abilities (e.g. \cite{O}). The working memory approach to mathematics, studies the roles of memory components in mental mathematical activities such as mental arithmetic and reasoning. Advanced brain scanning and imaging techniques provide available methodologies to explore associations between various mathematical activities and brain structural and functional areas. There is also some neural network modeling for mathematical cognition based on connectionist approach such as Parallel-Distributed Processing (e.g. McClelland et al. \cite{McMHYL}). For more related works on mathematical cognition we refer the readers to the book \cite{C}.

\noindent If we plan to explore the relations between mathematical thoughts and abilities in one side, and cognitive faculties and processes on the other side, we should notice the following distinction between two different aspects of mathematics. The first aspect considers mathematics as a crystallized formulated knowledge, consisting of all mathematical postulates, definitions and proved facts in the form of formalized topics or theories. This crystallized knowledge is clearly accumulated increasingly for each individual scholar of mathematics. But surprisingly it has not been initiated from vacuum. Humans have innate brain substrates by which are able to quantify the objects of environment, perceive differences in number, see numeric correspondences between different sets of objects; and perform elementary operations with numbers such as estimation of magnitude, counting and simple arithmetic. This competence known as Approximate Number System (ANS) (also known as Number Sense) is prior to any verbal education and exploits a visuo-spatial encoding for numbers \cite{PIPLD,D}. In addition to this competence, the brain is able to recognize and distinguish the fundamental patterns of Euclidean geometry such as congruency, connectedness, parallel lines and some other elementary metric and topology notions. This ability known as intuitive or core knowledge of geometry is innate and prior to education and culture \cite{DIPS,IPDHS}. Hence, the mathematical knowledge is based on basic mathematical substrates such as ANS and core knowledge of geometry. Although educated brains use purely symbolic concepts of numbers and operations as well as sophisticated theories of geometry but the basic substrates of mathematics are still used while mathematical thinking activities.

\begin{figure}[ht]
\begin{center}
\includegraphics[scale=0.5]{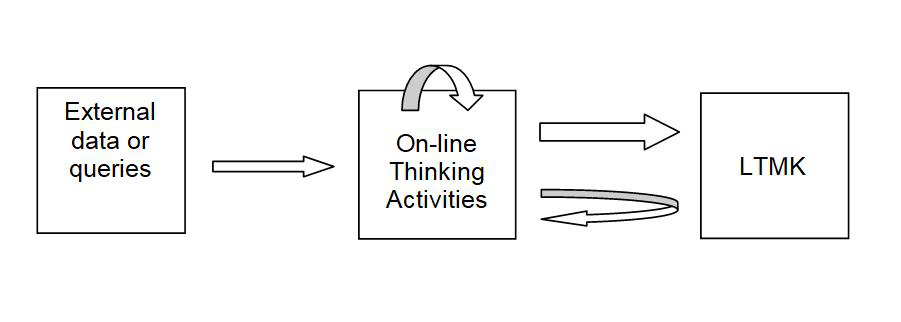}
\end{center}
\vspace*{-0.7cm}\caption{A diagram representing different aspects of mathematical thinking activities}\label{LTMK}
\end{figure}

\noindent The second aspect, considers mathematics as an on-line mental processing. This second aspect focuses on all mental processes which are involved while doing ongoing mathematical tasks such as calculation, proving a new result and mental logical inference. Sometimes further knowledge is obtained by reformulation or direct inference of previous knowledge. But basically, the formalized knowledge of mathematics is evolved through on-line mathematical activities. By LTMK (long term mathematical knowledge) we mean all previous knowledge of mathematical objects, theorems, techniques and skills stored in the brain. On-line thinking activities initiates either with some input from external challenges (e.g. someone asks you to do a calculation) or by activating some data and queries from LTMK.

\noindent Figure \ref{LTMK} depicts the relations between these components. It illustrates the framework of thought in the present paper. If necessary, the on-line component repeatedly retrieves information such as known facts and techniques from LTMK. In this case some information of LTMK is activated in the working level. In fact after processing in the on-line component, the obtained results transform into portions of LTMK in the form of final propositions. This stage is specified by the huge arrow in Figure \ref{LTMK}.

\noindent We make two comments concerning the above-mentioned distinctions between the two aspects of mathematics. The first is that when mathematicians evaluate their mathematical achievements and the question how their mind are led to obtain these findings, they usually judge by the final formalized form of the obtained results and simply overlook the vital role of the on-line part of mental activity in obtaining these results. It is not easy to judge what exactly is going on in the brain circuits during on-line processes and usually the mechanism of this stage is covert for mathematicians. It is instructive to mention a main instance in this regard. After achieving his mathematical theory for mechanics, Joseph-Louis Lagrange in a famous quotation declares:
``I have set myself the problem of reducing mechanics and the art of solving the problems appertaining to it, to general formulas, whose simple development gives all the equations necessary for the solution each problem ... No diagram will be found in this work. The methods which I expound in it, demand neither construction nor geometrical or mechanical reasoning, but solely algebraic operations subjected to a uniform and regular procedure." (quoted in \cite{Be}).
Now we explain some differences between the two aspects of mathematics. Corresponding to each of these aspects we have different levels of mathematical processing. When we want to evaluate mathematical knowledge of any individual we usually ask quick or yes/no questions. For example, what is the sum of angles in any convex quadrilateral? Is it true that any continuous function on a compact set is uniformly continuous? Assume that in a hypothetical task a mathematician is required to judge the validity of a mathematical statement $P$. In such a situation, it is not possible to perform the whole process of proving the statement $P$. Instead, he/she tries to recollect the required and related information from the categorized knowledge which logically leads to the validity of $P$. In other words, the brain tries to obtain some partial propositions related to the target proposition $P$ and relate them logically to $P$ and finally infer the validity of $P$. The questions corresponding to the second aspect is the ones which require on-line manipulating and processing of mathematical data. Moreover, the cognitive processes and functions involving in these aspects are different. For the first aspect, the normal tasks are ``semantic judgment" tasks, in which participants judge the validity of a mathematical statement. But for the second aspect, the normal tasks are on-line tasks in which on-line and transient data are manipulated and processed in the brain. An important point is that although direct inferences from LTMK and corollaries of previous theorems have some contributions in advancement of mathematics but the process of research in mathematics is mainly advanced by on-line mathematical activities. The cognitive systems corresponding to the second aspect is different from that of the first aspect in that the systems responsible for the on-line tasks (i.e. the second aspect of mathematics) is working memory components.

\noindent Our approach in this paper for investigating the involvements of verbal faculties in various mathematical processes, is to analysis the mathematical activities corresponding to the second aspect and investigate some related tasks. Some neuro-psychological theories suggest that they are founded upon evolutionarily ancient brain circuits for number and space, and others that they are grounded in language competence. The authors of \cite{AD} evaluate which brain systems underlie higher mathematics. They scanned professional mathematicians and mathematically naive subjects of equal academic standing as they (quickly) evaluated the truth of advanced mathematical and non-mathematical statements. They conclude that high-level mathematical expertise and basic number sense share common roots in a nonlinguistic brain circuit.  According to \cite{AD} their results suggest that high-level mathematical thinking makes minimal use of language areas and instead recruits circuits initially involved in space and number.  According to the distinctions we explained concerning the first and second aspects of mathematics, the type of tasks which have been used in the study of \cite{AD} corresponds to the first aspect of mathematics. We comment once again that research in mathematics is advanced by on-line mathematical activities. Hence the results of \cite{AD} are limited to some parts of mathematical thinking and it should not be concluded from their experimental results that general high-level mathematical thinking has almost nothing to do with language, even if the subjects are expert mathematicians.

\noindent {\bf The article is organized as follow.} The next section is devoted to review some related results and works. In Section 3 we introduce some key terminology such as mathematical thinking activity, data configuration, transform function, symbolic/formal processing, visual/spatial processing. We show how these mathematical processes can be distinguished using some mathematical tasks and dual-task methods. Section 4 discusses more topics about mathematical activities and visual/spatial processing. In Section 5 we itemize some spatial rehearsal mechanisms with related tasks and also explain the specific roles of language-based processes such as verbal articulation and inner instructions in various processes of mathematics.

\section{A memory theoretic introduction to mathematical cognition}

\noindent Complete description of the cognitive mechanisms involving in various mathematical tasks is very difficult. But many successful methodologies and consequent experimental achievements have been resulted. In this section we first briefly introduce the multi-component theory of working memory (M-WM). Initiated by the paper of Hitch in 1978 \cite{H}, many researchers have investigated involvement of the components of M-WM in various mathematical tasks such as mental arithmetic. Also, some authors study the roles of working memory capacities in academic achievement and math performance, for children and adults. In this section we review some related works in this direction and also some papers which have used brain scanning methodologies.

\noindent The working memory is a cognitive brain system which is responsible for active and on-line retention and rehearsal of information for mental activities. In addition to maintaining and recollection of information, the working memory is capable of processing information in a highly accessible level. Information may be entered to working memory either from external data or by activating information from long term memory (see Figure 1), but while being processed in working memory, these information are highly accessible and naturally transient. When a chunk of information is processed by the working memory, we say that the information is processed at the working level. The working memory includes short term memory in the sense that any mental processing of information which has short term nature can be described in terms of working memory. A theory for investigating working memory was presented by Baddeley and Hitch in 1974 \cite{BH} and has been completed twice by Baddeley. This completed theory is referred as the multi-component theory of working memory (hereafter M-WM). We refer the readers to the books \cite{B1,SK} for comprehensive explanations of the theory. It should be mentioned that the multi-component theory of working memory is not the only theory for working memory. There are other theories of working memory which emphasis on attention and capacity. But we employ the multi-component theory because this theory emphasis on the role of sub-systems and investigates specific modalities separately.

\noindent Now, we introduce briefly three components of M-WM. The phonological or articulatory loop component (PL) is thought to consist of two subcomponents. Phonological store is responsible for short term storage of verbal information in the form of a sound-based (phonological) code. Articulatory rehearsal is responsible for actively refreshing and elaborating information in the phonological store. In fact the rehearsal process in PL consists of subvocal articulation and is identified with verbal rote rehearsal. The contents of the passive store are thought to be maintained by subvocal rehearsal. PL has other functions to be explained later. The visuo-spatial sketchpad (VSSP) is the second component and is responsible for maintaining visuo-spatial information. The VSSP consists of two subcomponents: visual cache for storing the visual features and complexities of the objects and an active visuo-spatial rehearsal system, the so called inner scribe. It is generally believed that visuo-spatial working memory is also responsible for more active visuo-spatial tasks such as imagery. According to \cite{Z} visuo-spatial WM is responsible for tasks such as visual imagery, spatial rotation and spatial reasoning. The paper \cite{AJR} explains more involvement of visuo-spatial rehearsal in working memory. The visuo-spatial rehearsal and imagery are very important cognitive processes and significantly involved in mathematical thoughts. In the present paper we provide some mathematical tasks which are behaviorally associated to motor imagery.

\noindent M-WM also contains a regulating attentional system the so-called central executive (CE). The CE is an attentional control system which coordinates the processes of phonological and visuo-spatial information. The CE allocates attention to activities of PL and VSSP and is responsible for retrieval of information from long term memory. We explain in the next sections that task switching in mathematical activities and careful execution of procedures and algorithms demand executive control via inner speech in the component PL. It was also discovered that stress and anxiety use executive resources. For this reason anxiety reduces math performance of students \cite{AK}.
To the best of the author's knowledge, the first study with the aim of investigation of the role of working memory in mental arithmetic was done by Hitch \cite{H}. Hitch suggested that working memory made a crucial contribution to mental arithmetic. Logie and Baddeley in \cite{LB} investigated the effects of articulatory suppression and unattended speech on performance in simple counting tasks. Results were consistent in showing substantial disruption of counting performance by concurrent articulatory suppression. Much stronger results were obtained in \cite{LGW} concerning the significant role of phonological loop and subvocal articulation in counting and mental arithmetic. The experimental data of Logie et al. in 1994 \cite{LGW} support the view that the subvocal rehearsal component of working memory provides a means of maintaining accuracy in mental arithmetic. The authors of \cite{LGW} concluded that there was a minor involvement of the VSSP, which, furthermore, was restricted to the cases in which the calculations were presented visually (and could thus be restricted to a pre-calculation stage, such as in the encoding of the visual problem). The strong involvement of phonological loop in arithmetic was also verified in \cite{NDAS} in 2001, where the authors showed that the phonological loop plays a major role in mental addition, whereas the VSSP does not seem to be particularly involved. Thus, in mental computation, the storing space is verbal, and if calculations are presented visually in digits, a phonological recoding process first takes place.

\noindent As we mentioned in the introduction, some basic approximate counting and simple arithmetic use visuo-spatial representations of numbers and operations with numbers. However, when operations are more complex or required task switching and or precision, the visuo-spatial representation is not usable and hence the operation should be performed via symbolic/formal processing.
In addition to the strong involvement of working memory in mathematical processing, critical dependence of mathematics performance on working memory has shown by many authors (e.g. \cite{AK}). Low mathematics performance is not only because of low speed or quality of mathematical processing of a person. Instead, some preoccupations such as fear and anxiety work like a resource-demanding secondary task during accomplishment of main mathematical tasks and decrease significantly the performance. The authors of \cite{AK} examined the cognitive consequences of math anxiety and showed how performance on a standardized math achievement test varies as a function of math anxiety, and that math anxiety compromises the functioning of working memory. High performance of executive functions and lack of interruptive preoccupations are two effective factors in mathematical performance.

\section{Fundamental Mathematical Processes}

\noindent The aim of this section is to introduce three mathematical processes. These processes are extensively used in on-line mathematical activities. Their relationships with the working memory components will be shown. For this reason they are interpreted as fundamental mathematical processes. To begin, we need to introduce the following regulating concepts: mathematical thinking activity, data configuration and transform function. These concepts are explained with more details in Section 4. By a mathematical activity we mean any on-line mathematical thinking processing in which the goal is to accomplish a mathematical task such as calculation, numeric or symbolic computation, proof of an assertion, checking the validity of a computation or a proof and or exploration to discover a new idea. A mathematical activity is either run purely mental and completely executed in the brain or worked out on paper and executed via the interaction of the brain and paper. In the second case tools such as paper or blackboard play a supplementary memory workplace for the brain. We mentioned earlier that the capacity to maintain information in working memory stores is very limited. Working out mathematical activities on paper overcomes noticeably the capacity limitations of brain memory components. But as will be explained, since the mathematical activities have by definition on-line and transient nature, manipulation, rehearsal and rumination of information during these tasks require working memory resources and hence are limited to the working memory characteristics of the brain.

\noindent By a data configuration, we mean any concatenation of mathematical objects and their mathematical inter-relationships which are registered in memory stores during execution of a mathematical activity. A mathematical activity $P$ begins from an initial data configuration $S_{initial}$ and during the execution of the activity, a present data configuration is transformed into a next one. Hence, to initiate a mathematical activity one is required to create and set up an appropriate data configuration in his/her mind. Even this initial step is set up by inner instructive speech such as ``Suppose this ..." or ``Assume that ...". By a transform function we mean any kind of mathematical process corresponding to the activity $P$ which transforms one configuration of $P$ into another one. Data configurations concern the representation of existing data and their relations but transform functions manipulate the data and perform mathematical operations with these data as input. Hence, a mathematical activity is specified by its data configurations and intermediate transform functions. Note that at any step of execution of an activity, once a transform function $F$ is specified, then the further step of the task is taken only by operating $F$ on the underlying data configuration. But how these transform functions are determined is a completely different issue. We are now ready to introduce the promised processes.

\subsection{Symbolic/Formal Processing}

\noindent In symbolic/formal processing, data configurations are represented as phonological and verbal format. The process of information in symbolic/formal processing is done by verbal elaboration, inner speech and symbolic manipulation. A symbolic/formal object is encoded by its phonological name. The meaning of a symbolic object is decoded by its phonological content. The meaning of a formal expression is determined by verbal elaboration or articulation of the expression. This will be explained more latter. In terms of mathematics, meaning of any object is specified by its formal relationships with other objects. This meaning is specified by learning how it is used in formal relations. When we repeat - in the form of speech - the way of using a symbolic object in a formalism, the sense and meaning of the object becomes definite. The meaning of previously known symbols and formal relations as well as computing techniques are stored in LTMK. But in addition to these known notations and relations, some auxiliary notations are defined and used while proving mathematical assertions. For example consider the following generic phrases. Let $X$ be the set consisting of all rational numbers satisfying a property $R$; or ``let $r$ be the supremum value of all real numbers satisfying a property $R$". In these examples, the phrases ``the set consisting of all rational numbers satisfying a property $R$" and ``the supremum value of all real numbers satisfying a property $R$" are the defining expressions of the set $X$ and the real number $r$, respectively. In these definitions the notations are syncopated notations. The meaning of $X$ or $r$ is only specified by the vocal/subvocal articulation of their defining expressions. Any time a syncopated notation is required to be decoded and used in the proof, the subject has to decode its definition by full verbal articulation of its defining expression. The ``symbolic syncopation" is used throughout mathematics. In particular, in symbolic/formal processing, data configurations use syncopated symbolic notations. In symbolic/formal processing, the inter-relationships between the objects are formal mathematical relations. Here, the technique of ``formal syncopation" is used in order for abbreviation of formal relations. Formal syncopation is a very common methodology in whole mathematics. For example the sum of finitely many summands and infinitely many summands are denoted by $\sum$ and $\int$, respectively. Hence, lengthy and complicated computations with infinite sums and series are replaced by formal computations with integrals. Method of generating functions is completely formal syncopation technique used in combinatorial analysis. Functional transforms such as Laplace transform is a kind of formal syncopation.

\noindent As we mentioned earlier, some mathematical activities involving numbers use visuo-spatial representation of numbers and operations. But in many other operations with numbers, visuo-spatial representations are not applicable and instead phonological encoding and verbal elaboration are required to perform such tasks. In fact, since in exact counting and precise arithmetic we use verbal encoding of numbers and operations, then they are included in symbolic/formal processing. The following task is a generic example of symbolic/formal process. It is presented as a pilot study.

\noindent {\bf Task A.} Define a sequence $a_1, a_2, \ldots$ of integer numbers as follows. The first two elements of the sequence are $a_1=1$ and $a_2=2$. For each $n\geq 3$ define $a_n=2a_{n-1}-a_{n-2}$. Determine in your mind the value of $a_{6}$.

\noindent The author has performed Task $A$ under different conditions. In ordinary version, participants perform Task $A$ without doing any other action. In version $A_1$, the participants perform Task $A$ and simultaneously repeat an irrelevant word or speech. In version $A_1$, the irrelevant inner or external speech is a secondary task. In version $A_2$, the participant performs Task $A$ and simultaneously moves his/her hand regularly left and right (with uniform speed). In version $A_3$, the participant performs Task $A$ on paper with an irrelevant inner speech or external articulatory suppression as a secondary irrelevant task. Experimental results show that the performance in Tasks $A_1$ and $A_3$ is significantly worse than the performance in Task $A_2$. The main point is that although the brain is able to use visual representation of numbers to help phonological perform of computations, but any external speech or irrelevant subvocal articulation disrupts the whole task because the execution of task needs careful execution of the procedure and task switching, i.e. alternately multiply and subtract. Task $A$ and similar examples suggests that task switching and accuracy in symbolic/formal activities demand intensely phonological resources such as verbal self-instruction. Task $A$ has also an algorithmic aspect, because the subject should follow step by step an arithmetic procedure to obtain the final solution. The author has tested other algorithmic tasks. Even if you perform such tasks using pen and paper you use verbal encoding of numbers and operations and specially inner-speech instructions because you have to perform and control the steps of the task. The roles of phonological loop and inner-speech in general cognitive tasks requiring task switching and action control have been studied in Baddeley et al. \cite{BCA} in 2001 and in \cite{EM} in 2003. Interestingly, mathematical activities provide a big area of tasks which make strong relationships between verbal self-instruction and task switching as well as procedure control. Execution of algorithms, following flowcharts and computational procedures are some instances of these mathematical activities.

\subsection{Visual/Spatial Processing}

\noindent Data configurations in visual/spatial processing consist of pictures, shapes, volumes and other objects which are represented in visuo-spatial form such as knots and combinatorial graphs. This processing concerns the visual features and spatial relations of objects. The relations between objects are spatial such as geometric relations. Visual/spatial process includes visuo-spatial manipulations such as spatial movements, rotations, navigation etc. In addition to the above-mentioned realm, the visual/spatial processing has a minimum but efficient use in symbolic/formal processes. Symbols, equations and expressions are firstly represented visually and then their phonological contents will be used in symbolic/formal processes. This means that in terms of memory there should be a link between visual encoding of symbols such as algebraic variables and Arabic numerals and their phonological format. Another important point is that correct syntax of formal relations needs appropriate spatial relations between the symbols. Visual/spatial processing has another application in symbolic/formal operations which will be explained in the next section. It should be mentioned that when symbols and all kinds of relations are learned for the first time, then switch of encoding from visuo-spatial appearance of notations toward semantic and verbal contents, is necessary to decode the phonological meaning of syncopated information.

\noindent For mathematical visual/spatial processing the situation is completely different. The main difference is that spatial rehearsal and visuo-spatial imagery involving mathematical items such as geometric ones are major processes in this type of process. In some branches of mathematics such as projective and synthetic geometries we observe the striking role of visuo-spatial rehearsal and imagery, because these fields study figures and geometric objects as such, without recourse to coordinates and formulas. Importance of visuo-spatial imagery in branches of mathematics such as geometry has been insisted by many professional mathematicians (e.g. Conway et al. \cite{CDGT}). Also the role of visuo-spatial sketchpad in academic achievement in geometry has been studied in some papers (e.g. \cite{GMRC}). Although the mathematical descriptions of symbolic/formal and visual/spatial processes clearly differentiate their distinct natures, but cognitive distinction of these processes is not easily understood. Because even in simple symbolic/formal tasks such as mental computation the brain may use visual format of information. Using the so called dual-task technique, we propose a cognitive methodology to clarify the distinction between symbolic/formal and visual/spatial processing. In the following we present our tasks and experimental results. We note that the mathematical content of these tasks are very common in mathematical visual/spatial processes.

\begin{figure}[ht]
\begin{center}
\includegraphics[scale=0.6]{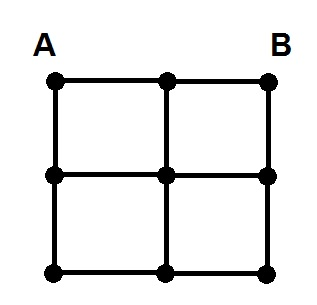}
\end{center}
\caption{A picture related to Task $B$}\label{pic task 2}
\end{figure}

\noindent {\bf Task B.} In this task, first the graph displayed in Figure \ref{pic task 2} is exposed to each individual subject so that its memory is fixed completely in his/her mind. Then in absence of the figure the experimenter is required to explore all paths from $A$ to $B$ which goes through each node exactly once and say the number of such paths.

\noindent The author has performed Task $B$ under different conditions. In the ordinary perform of Task $B$ the participant performs Task $B$ without any extra action. In version $B_1$, participants perform Task $B$ and at the same time repeat an unrelated word or sentence. In version $B_2$, the participant performs Task $B$ and at the same time moves his/her hand regularly left and right (with uniform speed). The experimental results show that irrelevant inner speech or external speech does not reduce performance of the participants, but irrelevant hand movement do. In fact the performance in $B_2$ is significantly reduced in contrast to the ordinary Task $B$. The results show that visual/spatial processes have close associations with visuo-spatial sketchpad and mental motor system but no dissociation with verbal component. Study of mental navigation as cognitive task goes back to 1968 by Brooks \cite{Br}. The relation between mental animation and visuo-spatial sketchpad has been studied in \cite{SH}. Imagery and spatial representation have significant roles in learning \cite{SchH}.
The author has performed other ``path construction mental tasks". In the following task which we call ``cube navigation task", experimenters are required to construct mentally a closed path.

\noindent {\bf Task $C$.} Imagine a cube in the three dimensional space. Find a closed path along the edges of the cube which goes exactly once through each corner.

\noindent Similar to Task $B$, the author has performed Task $C$ under different conditions. In the ordinary version Task $C$ is performed without any other action. In version $C_1$, irrelevant speech is a secondary task and in $C_2$, irrelevant hand movement is a secondary task. The results are all consistent with the results of Task $B$. The irrelevant inner speech or external speech does not reduce meaningfully performance of the participants, but irrelevant hand movement crucially decreases the performance. It should be mentioned that if we print 2-dimensional version of a cube on paper and perform Task $C$ on this figure in paper, we obtain similar results. It suggests that even when mathematical visual/spatial processes are performed using pen and paper, there exists close association with visuo-spatial scratchpad and motor imagery.

\subsection{Combinatory Processing}

\noindent Usually in mathematical tasks symbolic/formal and visual/spatial processes are combined in various ways. For example, assume that in a geometric picture including a number of lines and angles, the size of some angles are given. The task is to obtain the other angles. In this task we need to do some arithmetic but the corresponding quantities come from some visual information (i.e. the size of angles). The initial configuration consists of visuo-spatial and numerical information. In such tasks the processes are combined but the combination are separable. In general, the symbolic/formal and visual/spatial processes can be combined in separable ways in mathematical tasks. But as explained below, there exist many mathematical objects whose identity cannot be grasped by single or combination of the previous processes; they are neither symbolic/formal nor visual/spatial objects. Also, the relationships between the new objects cannot be reduced to single or combination of the two previous processes. In fact, the new objects are amalgamation of some visuo-spatial and symbolic information. We call such items ``bound objects" because some symbolic and visuo-spatial information bind together to form the bound objects. The identification of mathematical bound objects is determined by decoding the symbolic parts and visual/spatial relations between them, simultaneously. We present some examples of bound objects. A permutation on $5$ symbols such as $43521$ is a bound object. The simple task that counts the number of permutations on $5$ symbols is neither a symbolic/formal nor a visual/spatial processing. Permutations are composed of spatial arrangement of numbers from $1$ to $n$. The other example is labelled graphs. A graph is composed of a number of vertices on the plane and some edges between them. A (vertex) labelled graph is a graph whose vertices are labelled by some symbols such as numbers or letters. In on-line working with labelled graphs, the brain has to process the visual information of the graph and symbolic information of labels, simultaneously. There are also complex bound objects such as magic squares. Figure \ref{magic} displays an array of numbers which is called ``magic square". In the array, the numeric sum of entries in each row, each column and principal diagonals of the array are a same number, i.e. 175 (in this example). These conditions define the concept of magic squares. To specify the identity of a magic square, one needs to determine the sum of some numbers which are arranged in certain spatial locations. The arrangements along a row, a column or a diagonal are these spatial arrangements. Note that the precise summation of a few integers is in general a verbal operation. Moreover, only the sum of numbers which are placed in some certain spatial arrangements, are required to make the identity of magic squares. Hence, magic squares are bound objects. Figure \ref{magic} illustrates a magic square.

\begin{figure}[ht]
	{\Large
\begin{center}
\begin{tabular}{|c|c|c|c|c|c|c|}
\hline22&47&16&41&10&35&4\\[.4eM]
%\hline&&&&&&\\[-.9eM]
\hline5&23&48&17&42&11&29\\[.2eM]
\hline&&&&&&\\[-.9eM]
30&6&24&49&18&36&12\\[.2eM]
\hline&&&&&&\\[-.9eM]
13&31&7&25&43&19&37\\[.2eM]
\hline&&&&&&\\[-.9eM]
38&14&32&1&26&44&20\\[.2eM]
\hline&&&&&&\\[-.9eM]
21&39&8&33&2&27&45\\[.2eM]
\hline&&&&&&\\[-.9eM]
46&15&40&9&34&3&28\\[.2eM]
\hline
\end{tabular}
\end{center}
}\caption{A magic square}\label{magic}
\end{figure}

\noindent The other complex bound objects are very well-known in mathematics. By an $n\times n$ Latin square $L$ we mean any square array consisting of the entries $1, 2, \ldots , n$ such that each number is appeared exactly once in each row and each column of $L$. In fact, in bound objects we find visuo-spatial patterns consisting of symbols in which the symbols form some spatial arrangements. Combinatorial analysis is a branch of mathematics which deals with objects such as permutations, Latin squares, labelled graphs and other similar objects which according to our definition are bound or amalgamated objects. For this reason we term the new process as combinatory processing. A combinatory processing is a mathematical on-line processing which works with bound objects. The mathematical relationships between bound objects are not reduced to any combination of visual/spatial and symbolic/formal processes. Of course it is possible to construct a formalism between appropriate bound objects. In this case, the objects are bound objects but the relations are formal, i.e. the combination of symbolic/formal and combinatory processing. To the best of author's knowledge, cognitive study of tasks concerning mathematical bound objects has not been done yet. This is an unexplored area. We know almost nothing about the ways the brain employs to process the combinatory objects.

\section{Mathematical activities and processes}

\noindent Recall that by mathematical activity we mean any on-line mathematical thinking processing in which a mathematical task is accomplished. Mathematical activities consist of data configurations and corresponding transform functions. Data configurations are stimulated by external stimuli such as questions or through activation of some information from long term mathematical knowledge (LTMK). During mathematical activities, in addition to the initial phase or configuration, some new data and information - which form the further data configurations of the activity - are produced in mind which are maintained and represented by memory stores. In many mathematical activities which we related to the second aspect, the information under process should be highly active and accessible, and hence transient. Moreover, the exploration of ideas and partial solutions during solving mathematical problems, needs processing of transient but active mathematical information in mind. We conclude that the mathematical activities are mainly accomplished in working level. Storage of data configurations requires working memory stores and process of transform functions requires working memory mechanisms such as rehearsal and articulation. In fact, working memory serves like a scratchboard in the brain for on-line mathematical activities. Recall Figure \ref{LTMK} which depicts connections between these components. Of course there are mathematical tasks which are done by routine recollection of previous knowledge from LTMK. We related such tasks to the first aspect of mathematics and they happen when we are asked to judge or recollect quickly the validity of a mathematical statement. There is no need to make an exact demarcation between the tasks related to the first and second aspects of mathematics. But there are salient differences between cognitive resources and processes regarding to the aspects which we explained.

\begin{figure}[ht]
\begin{center}
\includegraphics[scale=0.5]{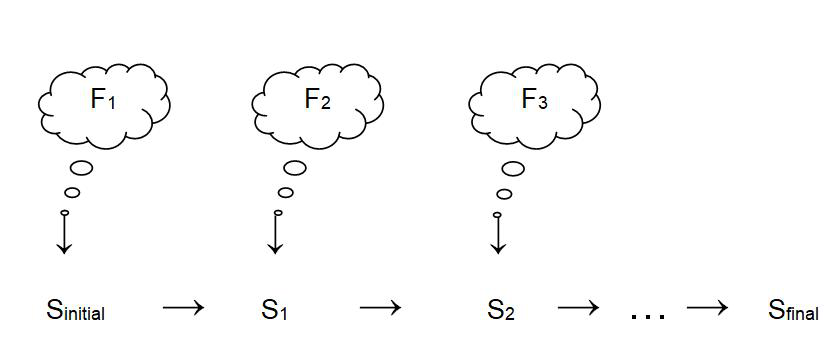}
\end{center}
\caption{Configurations and transform functions}\label{ini}
\end{figure}

\noindent Note that at any step of execution of an activity, once a transform function say $F$ is determined, then the further step of the task is taken only by operating $F$ on the underlying configuration. But how these transform functions are determined is a completely different issue. Here we confront with a dichotomy. Checking the validity of a computation or a proof and in general, operating a definite function on an input data configuration, are in one side of the dichotomy. Production of an algorithm or a proof and in general, search and determination of a transform function are in the other side of the dichotomy. Therefore, corresponding to any mathematical activity there exits an accompanying mathematical processing which explores and determines all transform functions required for the accomplishment of the activity. This accompanying process is in fact a logical reasoning which guides the whole thinking activity toward the final step of the task. This inner reasoning is done by inner speech which directs and controls the thought. The following helpful figure illustrates the fact that the execution and discovery of transform functions belong to different categories. In Figure \ref{ini} the process begins by an initial data configuration $S_{initial}$ and is progressed via intermediate configurations $S_1, S_2, \ldots$ to attain the final situation. At each step $S_i$, an appropriate transform function $F_{i+1}$ is determined and the next step $S_{i+1}$ is obtained by operating $F_{i+1}$ on the configuration $S_i$.

\noindent Of course in general, evolution of configurations in mathematical activities is a continuum of mental mathematical configurations. We don't need to discretize it. In Figure \ref{ini} they are presented in sequential form just for simplicity. It is possible that in an activity, some transform functions are symbolic/formal and some visual/spatial. Visual/spatial processing has at least two essential roles for symbolic/formal thinking. We have mentioned before the first role. Correct logical and mathematical syntaxes of all formal relations require appropriate spatial concatenation of the numeric and functional symbols and logical connectors. Hence, checking the validity of syntaxes is based on correct spatial placements of symbols with respect to each other. The second role is effective and much more non-trivial than the first role. Assume that in a mathematical activity we have a sequence of symbolic/formal configurations. The search for transform functions sometimes needs suitable rearrangements of symbols and expressions. New arrangements may help mind find new and effective formal expressions and relations between symbols. Consider a task in which the sum of numbers from $1$ to $100$ is required with the minimum possible number of arithmetic operations. One representation is $1+2+ \cdots +100$, for which $99$ sum operations should be done. Now consider the following arrangement

\begin{center}
\begin{tabular}{ccccccc}
1&2&3&4&$\cdots$&99&100\\[.4eM]
%\hline&&&&&&\\[-.9eM]
100&99&$\cdots$&4&3&2&1\\[.2eM]
%\hline&&&&&&\\[-.9eM]
\end{tabular}
\end{center}

\noindent By this configuration, the number of operations is only 3. Because all pairs in each column has a same sum $101$. Hence the total sum is $100 \times 101$. But we need the half of it, i.e. $5050$. An equivalent task is to determine the following sum $1+2+ \cdots +100+1+2+ \cdots +100$. The point is that the suitable pairing of this list is achieved by visuo-spatial scanning of whole data. Some other examples of this sort, are putting parentheses in suitable places in long expressions involving arithmetic or algebraic operations in order for simplify the whole expression or cancel some terms. In general, spatial rearrangements and visual scanning of formal expressions are some key roles of visuo-spatial process in symbolic/formal activities. Sometimes mathematicians overlook these key roles because in some cases such visuo-spatial processes are done completely mentally and transient in the mind of the mathematician.

\subsection{On language and intuition in mathematics}

\noindent What are the roles of language and intuition in mathematical cognition and which one has the most significant role? This question attracted many mathematicians and philosophers from long time ago. For example disputes between intuitionists and formalists on foundation of mathematics have roots in this question. This question in this form seems too simplified, because we have many aspects of language and intuition involved in mathematical thoughts and abilities. In this section we describe the roles of language and intuition in a memory theoretic framework. By visuo-spatial intuition (or simply intuition) we mean the mental faculty which exploits the wide abilities of visual/spatial processing in mathematical activities including spatial representations and transform functions. Spatial rehearsal, in general speaking, is interpreted as active and dynamic processing of visuo-spatial information. Hence, visuo-spatial imagery and spatial rehearsal are the main cognitive characteristics of visuo-spatial intuition. We have already insisted the use of visual/spatial process even in symbolic/formal activities in terms of visual scanning and spatial rearrangement of formal expressions. Even in advanced levels of mathematics, on-line activities, in particular of visual/spatial type, recruit working memory resources such as visuo-spatial representations, spatial rehearsal, mental movements and imagery. In Section 3 we presented some tasks which are typically visuo-spatial. Two similar tasks of type ``mental path construction" are presented in the following. Note that these tasks like the previous ones are not artificial tasks but they are extracted from real mathematical problems. Various spatial rehearsal and imagery techniques are required to accomplish these tasks.

\noindent {\bf Task 1.} Take a $3 \times 3 \times 3$ array of dots in space, and connect them by edges up-and-down, left-and-right, and forward-and-back. The task is to try and answer does there exists a closed path which visits every dot exactly once?

\noindent {\bf Task 2.} Do Task 1 for a $4 \times 4 \times 4$ array of dots, finding a closed path that visits every dot exactly once.

\noindent The following tasks need different visuo-spatial rehearsal techniques.

\noindent {\bf Task 3.} A tetrahedron is a pyramid with a triangular base. Rest a tetrahedron so that it is balanced on one edge. What shape is the topmost part of the tetrahedron in this situation? Now, in your mind slice the tetrahedron horizontally halfway between its lowest edge and its topmost part. What shape is the slice?

\noindent {\bf Task 4.} An octahedron is the shape formed by gluing together equilateral triangles four to a vertex. Rest an octahedron on a face. What shape is the topmost part of it? Now, slice the tetrahedron halfway up. What shape is the slice?

\noindent {\bf Task 5.} How many colors are required to color the faces of an octahedron so that faces which share an edge have different colors?

\noindent The following visuo-spatial task is an example of dissection problems in geometry. It is an elementary form of higher geometric operations on surfaces the so called geometric surgery. Dissection and surgery are mainly mental and imagery processes. The following has been chosen from plenty of problems.

\begin{figure}[ht]
\begin{center}
\includegraphics[scale=0.50]{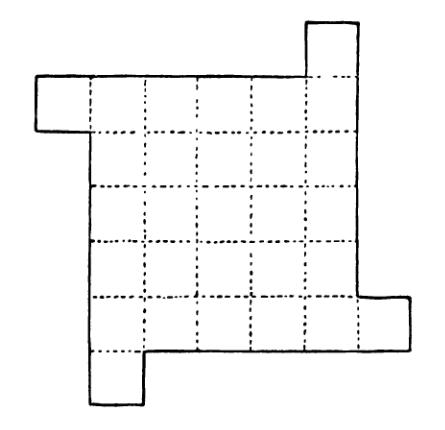}
\end{center}
\caption{The figure of Task 6}\label{task6}
\end{figure}

\noindent {\bf Task 6.} Cut the figure displayed in Figure \ref{task6} into four pieces that will fit together and form a square.

\noindent Finally, we itemize some of the known visuo-spatial rehearsal mechanisms in mathematical activities.

\begin{itemize}
\item{Focus and attention at some part or some features in a data configuration (e.g. Tasks 3, 4 and plenty of other examples)}
\item{Grasp of a panoramic or particular perspective (e.g. Task 5)}
\item{Navigation, routing, path construction and in general mental movements (e.g. Tasks B, C and Tasks 1, 2)}
\item{Cutting and pasting geometric pieces, dissection and surgery of geometric figures (e.g. Task 6)}
\item{Determination or estimation of distances and magnitudes}
\end{itemize}

\noindent In the rest of the article by summarizing the previous explains, we specify the key roles of various aspects of language for mathematics. In Dehaene et al. 1999 \cite{DSPST} using behavioral tasks and fMRI analysis, the roles of language and intuition has been investigated. It was shown that approximate arithmetic shows language independence and relies on visuo-spatial processing. In contrast exact arithmetic is acquired in a language-specific format and employs word-association processes. From the viewpoint of the present article, tasks consisting of many partial sub-tasks (e.g. Task A) which require task switching and control of action, recruit language-specific representation and operation of numbers as well as inner speech to direct the execution and control of the procedure. In fact, the result of \cite{DSPST} is generalized as follows. Not only exact arithmetic but also all algorithmic tasks including computational ones are acquired by language-specific processes. Overall, we summary that in on-line symbolic/formal mathematical activities even using pen and paper, and even in advanced levels, phonological and verbal resources are intensively demanding. We list them as follow.

\begin{itemize}
\item{Verbal encoding and representation of numbers or abstract symbols}
\item{Decoding symbolic syncopations and formal syncopations using verbal articulation}
\item{Task switching and control of accuracy via inner speech}
\item{Correct execution and control of procedures according to flowcharts and algorithms}
\item{Mental logical reasoning and inference}
\end{itemize}

%%%%%%%%%%%%%%%%%%%%%%%%%%%%%%%%%%%%%%%%%%%%%%%%%%%%%%%%%%%%%%%%%%%%%%%%%%%%%%%%%


\begin{thebibliography}{99}

\bibitem{AD}
M. Amarlic, S. Dehaene, Origins of the brain networks for advanced mathematics in expert mathematicians,  Proceedings of the National Academy of Sciences, 113 (2016) 4909--4917.

\bibitem{AK}
M. H. Ashcraft, J. A. Krause, Working memory, math performance, and math anxiety, Psychonomic Bulletin $\&$ Review, 14 (2) (2007) 243--248.

\bibitem{AJR}
E. Awe, J. Jonides, P. A. Reuter-Lorentz, Rehearsal in spatial working memory, Journal of Experimental Psychology: Human Perception and Performance, 24 (1998) 780--790.

\bibitem{B1}
A. D. Baddeley, Working Memory, Thought and Action, Oxford University Press (2007).

\bibitem{BCA}
A. D. Baddeley, D. Chincotta, A. Adlam, Working memory and the control of action: evidence from task switching, Journal of Experimental Psychology: General, 130 (2001) 641--657.

\bibitem{BH}
A.D. Baddeley, G.J. Hitch, Working memory, In: The Psychology of Learning and Motivation: Advances in Research and Theory, ed. G. A. Bower, pp 47--89, New York: Academic.

\bibitem{BLV}
A. D. Baddeley, V. Lewis, G. Vallar, Exploring the articulatory loop, Q. J. Exp. Psychol. A 36 (1984) 233--252.

\bibitem{Be}
E. T. Bell, The development of mathematics, Dover (1992).

\bibitem{Br}
L. R. Brooks, Spatial and verbal components of the act of recall, Canadian Journal of Psychology, 22 (1968) 349--368.

\bibitem{C}
J. I. Campbell (Ed.), Handbook of mathematical cognition, Psychology Press (2005).

\bibitem{CDGT}
J. Conway, P. Doyle, J. Gilman, B. Thurston, Geometry and the imagination,
Based on materials from the course taught at the University of Minnesota Geometry
Center in June 1991, Version 0.941 (2010).

\bibitem{D}
S. Dehaene, The number sense: How the mind creates mathematics, Oxford University Press (2011).

\bibitem{DIPS}
S. Dehaene, V. Izard, P. Pica, E. Spelke, Core knowledge of geometry in an Amazonian indigene group, Science 311 (2006) 381--384.

\bibitem{DSPST}
S. Dehaene, E. Spelke, P. Pinel, R. Stanescu, S. Tsivkin, Sources of Mathematical Thinking: Behavioral and Brain-Imaging Evidence, Science 284 (1999) 970--974.

\bibitem{EM}
M. J. Emerson, A. Miyake, The role of inner speech in task switching: a dual-task investigation, J. Mem. Lang. 48 (2003) 148--168.

\bibitem{F}
S. Feferman, Mathematical intuition vs. mathematical monsters, Synthese 125 (2000) 317--332.

\bibitem{GMRC}
D. Giofre, I. C. Mammarella, L. Ronconi, C. Cornoldi, Visuospatial working memory in intuitive geometry, and in academic achievement in geometry, Learning and Individual Differences, 23 (2013) 114--122.

\bibitem{H}
G. J. Hitch, The role of short-term working memory in mental
arithmetic, Cognitive Psychology 10 (1978) 302--323.

\bibitem{K}
I. Kant, Critique of Pure Reason, Cambridge University Press (1998) (translated and edited by Paul Guyer and Allen W. Wood).

\bibitem{IPDHS}
V. Izard, P. Pica, S. Dehaene, D. Hinchey, E. Spelke, Geometry as a universal mental construction, In: Space, Time and Number in the Brain, ed. S. Dehaene and E. Brannon, pp. 319--332, Academic Press (2011).

\bibitem{LB}
R. H. Logie, A. D. Baddeley, Cognitive Processes in Counting, Journal of Experimental Psychology: Learning, Memory, and Cognition, 1987, Vol. 13, No. 2, 310--326.

\bibitem{LGW}
R. H. Logie, K. J. Gilhooly, V. Wynn, Counting on Working Memory in Arithmetic Problem Solving, Memory and Cognition, 22 (1994) 395--410.

\bibitem{McMHYL}
J. L. McClelland, K. Mickey, S. Hansen, A. Yuan, Q. Lu, A Parallel-Distributed Processing Approach to Mathematical Cognition (2016).

\bibitem{NDAS}
M-P. No\"el, M. D\'esert, A. Auburn, X. Seron, Involvement of short-term memory in complex mental calculation, Memory $\&$ Cognition 29 (2001) 34--42.

\bibitem{O}
B. Ojose, Applying Piaget's theory of cognitive development to mathematics instruction, The Mathematics Educator, 18 (1) (2008).

\bibitem{PIPLD}
M. Piazza, V. Izard, P. Pinel, D. Le Bihan, S. Dehaene, Tuning curves for approximate numerosity in the human intraparietal sulcus, Neuron, 44 (3) (2004) 547--555.

\bibitem{P}
H. Poincar\'{e}, The value of science, New York, Dover (1958) (translated by George Bruce Halsted).

\bibitem{R}
B. Russell, The principles of mathematics, Cambridge Univ. Press (1903).

\bibitem{SchH}
D. L. Schwartz, J. Heiser, Spatial representations and imagery in learning (pp. 283-298), Cambridge Univ. Press (2006).

\bibitem{SH}
V. K. Sims, M. Hegarty, Mental animation in the visuospatial sketchpad: Evidence from dual-task studies. Memory \& Cognition, 25 (1997) 321--333.

\bibitem{SK}
E. E. Smith, S.M. Kosslyn, Cognitive Psychology, Prentice-Hall, Inc. New Jersey, USA (2007).

\bibitem{Z}
H. D. Zimmer, Visual and spatial working memory: From boxes to networks, Nuerosciences and Biobehavioral Reviews, 32 (2008) 1373--1395.

\end{thebibliography}
\end{document}